\documentclass[pdflatex,sn-mathphys-num]{sn-jnl}

\usepackage{array} 
\usepackage{multirow} 
\usepackage{hhline}
\usepackage{graphicx} 
\usepackage{amsmath,amssymb,amsfonts}%
\usepackage{amsthm}%
\usepackage{mathrsfs}%
\usepackage[title]{appendix}%
\usepackage{xcolor}%
\usepackage{textcomp}%
\usepackage{manyfoot}%
\usepackage{booktabs}%
\usepackage{algorithm}%
\usepackage{algorithmicx}%
\usepackage{algpseudocode}%
\usepackage{listings}%

\usepackage{caption}
\usepackage[labelformat=simple]{subcaption}

\DeclareCaptionLabelFormat{subcaptionlabel}{\normalfont(\textbf{#2}\normalfont)}
\usepackage{amsmath}
\usepackage[version=4]{mhchem}
\usepackage{siunitx}
\usepackage{longtable,tabularx}
\usepackage{color,soul}

\usepackage[nameinlink]{cleveref}
\usepackage{rotating}
\usepackage{cancel,soul,ulem}
\usepackage{booktabs}

\begin{document}

\title[Effects of turbulence boundary conditions on Spalart-Allmaras RANS simulations for active flow control applications]{Effects of turbulence boundary conditions on Spalart-Allmaras RANS simulations for active flow control applications}


\author[1]{\fnm{Navid} \sur{Monshi Tousi}}\email{navid.monshi.tousi@upc.edu}

\author*[1]{\fnm{Josep M.} \sur{Bergad\`a}}\email{josep.m.bergada@upc.edu}

\author[2]{\fnm{Fernando} \sur{Mellibovsky}}\email{fernando.mellibovsky@upc.edu}

\affil[1]{\orgdiv{Department of Fluids Mechanics}, \orgname{Universitat Polit\`ecnica de Catalunya}, \orgaddress{ \city{Barcelona}, \postcode{08034}, \country{Spain}}}

\affil[2]{\orgdiv{Department of Physics, Aeronautics Division}, \orgname{Universitat Polit\`ecnica de Catalunya}, \orgaddress{ \city{Barcelona}, \postcode{08034}, \country{Spain}}}


\abstract{We assess the suitability of Reynolds-Averaged Navier-Stokes (RANS) simulation using the Spallart-Almaras (SA) turbulence model as a closure in analysing the performance of fluidic Active Flow Control (AFC) applications. In particular, we focus on the optimal set of actuation parameters found by \citet{Tousi2021,tousi2022large} for a SD7003 airfoil at a Reynolds number $Re=6\times10^4$ and post-stall angle of attack $\alpha=14^\circ$ fitted with a Synthetic Jet Actuator (SJA). The Large Eddy Simulation (LES) presented in that work is taken as the reference to identify the best choice of boundary conditions for the turbulence field $\tilde{\nu}$ at both domain inlet and jet orifice in two-dimensional RANS-SA computations.
Although SA-RANS is far less accurate than LES, our findings show that it can still predict macroscopic aggregates such as lift and drag coefficients quite statisfactorily and at a much lower computational cost, provided that turbulence levels of the actuator jet are set to a realistic value. An adequate value of $\tilde{\nu}$ is instrumental in capturing the correct flow behaviour of the reattached boundary layers for close-to-optimal actuated cases.
This validates the use of RANS-SA as a reliable and cost-effective simulation method for the preliminary optimisation of SJA parameters in AFC applications.}





\maketitle

\section{Introduction} \label{sec:intro}
The application of Active Flow Control (AFC) technology to airfoils and bluff-bodies has usually the object of postponing or even suppressing flow separation thereby improving aerodynamic performance. AFC techniques fall into three main different categories \cite{Cattafesta_2011}, namely moving body, plasma and fluidic actuation. Moving body actuators act on the geometry of the body to inject momentum into the flow \citep{Wang_2019}. Plasma actuators generate fast temporal response jets of ionized fluid by applying large electric potential differences \citep{Cho_2011,Foshat_2020,Benard_2014,Benard_2016, de2020influence}. Finally, fluidic actuators (FA) inject/suck fluid to/from the boundary layer. These latter are, by far, the most common sort of actuators. Some recent advances towards the understanding of some such devices can be found in \citep{Baghaei_2019, Baghaei_2020, bergada2021fluidic}.

Among fluidic actuators, Synthetic Jet Actuators (SJA), also called Zero Net Mass Flow Actuators (ZNMFA), are often used due to their outright simplicity and demonstrated capability of suppressing flow separation \citep{glezer2002synthetic,rumsey2004summary,wygnanski2004variables,findanis2008interaction,zhiyong2020modulation}. SJAs do not require an external fluid supply, as the jet can be simply produced with an oscillating membrane housed inside a tiny cavity located just beneath the surface. SJAs have been shown more effective than Continuous Jet Actuators (CJA), at comparable power input levels, in improving the performances of a stator compressor cascade \citep{de2015comparison,traficante2016flow,de2012active,zhang2019comparison, amitay2001aerodynamic}.

Most research on SJAs has traditionally focused on the combined effect of just two parameters: the actuation frequency $f_j$ and the jet momentum coefficient $C_{\mu}$. The former is nondimensionalised with the airfoil chord $C$ and the free-stream velocity $U_\infty$ following $f_j C/U_{\infty}$, while the momentum coefficient is defined as $C_{\mu}={(\rho_{j} {U}_{j}^{2} h_j \sin\theta_j})/{(\rho_{\infty}U_{\infty}^{2}C)}$, with $h_j$ the jet width, $\rho_{j}$ and $\rho_{\infty}$ the jet and far field fluid densities, ${U}_{j}$ the maximum jet velocity and $\theta_j$ the jet injection angle with respect to the airfoil surface.

Among the first experimental studies on the effects of the momentum coefficient, jet frequency and jet position of a SJA, \citet{amitay2001aerodynamic,amitay2002role} focused on improving the performances of symmetric airfoils. They observed that placing the actuator close to the boundary layer separation point minimised the momentum coefficient required for flow reattachment. Furthermore, actuating with frequencies of the same order of magnitude as the natural vortex shedding frequency produced only unsteady reattachment, while full flow reattachment could be achieved by actuating at about ten times the vortex-shedding frequency.

Experimental and numerical studies on a SJA-actuated NACA0015 airfoil at $Re=8.96\times10^{5}$ by \citet{gilarranz2005new} and \citet{you2008active} followed shortly after. The effectiveness of the actuation was shown to be rather poor at angles of attack below $\alpha\leq10^\circ$, but greatly improved for $\alpha\gtrsim12^\circ$. For much higher post-stall angles of attack $\alpha>25^{\circ}$ the actuation frequencies required to obtain high lift coefficients were particularly large. Employing $C_{\mu}=0.0123$, $f_j=1.284 U_\infty/C$ and $\theta=30.2^{\circ}$, the lift increase with respect to baseline case obtained by the numerical simulations was $70\%$.
The same airfoil but at a lower Reynolds number $Re=3.9\times10^{4}$ was experimentally studied by \citet{tuck2008separation}, and simulated numerically with LES by \citet{kitsios2011coherent}. Maximum efficiency was obtained for SJA nondimensional frequencies around $f_j=1.3 U_\infty/C$ in the experiments. 
Numerical simulations revealed that the optimal frequencies coincided with the baseline vortex shedding frequency ($f_{\mathrm{vK}}$) and its first harmonic ($2\,f_{\mathrm{vK}}$). Particle Image Velocimetry (PIV) later confirmed these results \cite{buchmann2013influence}. \citet{itsariyapinyo2018large} studied yet the same airfoil using LES at $Re=1.1\times10^{5}$. The synthetic jet actuator was placed tangentially to the surface at the trailing edge of the airfoil. A linear relation was observed between the momentum coefficient of the jet and the resulting lift coefficient of the airfoil. The effect, however, saturated at a threshold, beyond which further increases of the momentum coefficient produced no net gain in the lift coefficient.

The Reynolds-averaged Navier-Stokes (RANS) equations with the k-$\omega$ SST model as their turbulence closure was applied by \citet{kim2009separation} on a NACA 23012 airfoil at $Re=2.19\times10^{6}$ and $\alpha\in[6^\circ,22^\circ]$ to assess flow separation control in 5 different slot/flap/jet configurations. Low actuation frequencies and large momentum coefficients were found most effective in reattaching the boundary layer in large separated regions. They also investigated multi-array / multi-location SJA implementations that successfully decreased the required effective jet velocity magnitude. The same airfoil and Reynolds number were studied by \citet{monir2014tangential} employing RANS with the Spallart-Almaras (SA) turbulence model. Aerodynamic efficiency was substantially increased with oblique SJA actuation at $\theta_j=43^\circ$, but tangential actuation was by far the optimum.
The effect of SJA on a NACA0025 airfoil at $Re=10^{5}$ and $\alpha=5^\circ$ was experimentally studied by \citet{goodfellow2013momentum}. They realized that the momentum coefficient was the primary flow control parameter, obtaining up to $50\%$ drag decrease for $C_\mu$ above a certain threshold. The same airfoil and Reynolds number were respectively assessed by \citet{feero2015flow, feero2017influence} at $\alpha=10^\circ$ and $\alpha=12^\circ$, respectively. Excitation frequencies around the vortex shedding frequency required momentum coefficients one order of magnitude lower than at higher frequencies to enforce reattachment. Flow control was more effective with the jet placed in the vicinity of the boundary layer average separation point for the baseline case. The effect of AFC excitation frequency on a NACA0018 airfoil at $\alpha=10^\circ$ and $Re=1000$ using three-dimensional Direct Numerical Simulation (DNS) was investigated by \citet{zhang2015direct}. Three different frequencies were assessed ($f_j=0.5 U_\infty/C$, 1 and 4), being $f_j=1 U_\infty/C$ the optimal one.

The Selig-Donovan 7003 (SD7003) \citep{selig1989airfoils,selig1995summary} airfoil was designed for low Reynolds number applications, and has been thoroughly studied at $Re=6\times 10^{4}$ via LES \citep{breuer2018effect,qin2018large,rodriguez2020effects}. \citet{breuer2018effect} tested several values of the turbulence intensity at domain inlet for $\alpha=4^\circ$, ranging from nil to $Tu=11\%$, with the aim of understanding the impact of free stream turbulence on the Laminar Separation Bubble (LSB). High free-stream turbulence levels were shown to reduce and even suppress the LSB, thereby enhancing aerodynamic performance. Time-dependent velocity oscillations of various amplitudes and frequencies were employed at domain inlet by \citet{qin2018large} at the same $\alpha=4^\circ$. Flow separation was delayed during the acceleration phase and advanced in the deceleration phase.
\citet{rodriguez2020effects} performed LES simulations at $\alpha=\{4^\circ,11^\circ$,$14^\circ\}$ with and without SJA. Aerodynamic efficiency was seen to increase by ${\Delta \eta/\eta}=124\%$ at $\alpha=14^\circ$, while actuation effectiveness largely decreased at pre-stall angles of attack.
  
Recent research in AFC application to airfoils has been extensive, with notable contributions made by \citet{Tousi2021, tousi2022large}. \citet{Tousi2021} undertook the simultaneous optimization of five SJA parameters on a SD7003 airfoil at a Reynolds number of $6\times10^{4}$ and various pre- and post-stall angles of attack ($\alpha={4^\circ,6^\circ,8^\circ,14^\circ}$) using a Genetic Algorithm (GA) and the RANS-Spalart-Allmaras turbulence model.

The key difference between \citet{tadjfar2020optimization} and \citet{Tousi2021} resides in the methodology employed. While the former employed an Artificial Neural Network (ANN) to optimise AFC parameters, the latter relied on far more accurate CFD simulations along the entire optimization process, which results in more reliable results but at the expense of increased computational cost. The simpler approach of straightforward parametric exploration \citep{couto2022aerodynamic} becomes unfeasible for the accurate optimisation of a large number of parameters.

The optimal SJA results of \citet{Tousi2021}, obtained with RANS-SA, were tested against LES by \citep{tousi2022large}. Although the LES runs achieved aerodynamic performances comparable to those predicted by the original RANS-SA optimisation, there appeared to be some disagreement both in mean flow topology and performance parameters. The discrepancies were traced back to a high sensitivity of RANS-SA results to the values of the modified turbulence viscosity $\tilde{\nu}$ at the velocity inlet boundaries of the domain, particularly in situations of reattached flow, as is mostly the case for AFC applicattions. Our primary focus here is to examine the impact of $\tilde{\nu}$ values at domain inlet and jet orifice, with the aim of providing recommendations as to how RANS-SA computations must be set up for AFC optimisation applications

The paper is structured as follows. The governing equations for the RANS-SA model are presented in \S\ref{sec:governingEquations}. Section \S\ref{sec:numericalModelling} discusses the computational domain, the mesh and the boundary conditions. The main results are then presented in section \S\ref{sec:results3} and conclusions drawn in \S\ref{sec:conclusions3}.

\section{Governing equations}\label{sec:governingEquations}

The original CFD simulations, performed in a previous optimization study \cite{Tousi2021}, were all based on the Unsteady Reynolds-Averaged Navier-Stokes (URANS) equations with the Spalart-Allmaras (SA) turbulence model \cite{spalart1992one} used as closure. The flow was considered incompressible. Employing Direct Numerical simulations (DNS) or Large Eddy Simulations (LES) were discarded due to the massive computational power requirements, which render these methods inapplicable to multi-parameter optimistion. The same turbulence model is employed here. The URANS equations read
\begin{eqnarray}
\frac{\partial {u}_{i}}{\partial x_{i}} & = & 0\\
\frac{\partial {u}_{i}}{\partial t}+{u}_{j}\frac{\partial {u}_{i}}{\partial x_{j}} & = & -\frac{1}{\rho} \frac{\partial {p}}{\partial x_{i}}+\frac{\partial}{\partial x_j} \left(\nu \frac{\partial {u}_{i}}{\partial x_{j}}-\overline{u_{i}^{\prime} u_{j}^{\prime}}\right)
\end{eqnarray}
where $u_i=(u_1,u_2,u_3)$ and $p$ are the Reynolds-averaged velocity components and pressure, respectively, $\rho$ and $\nu$ the density and kinematic viscosity of the fluid, and $R_{ij}\equiv\overline{u_{i}^{\prime} u_{j}^{\prime}}$ the components of the Reynolds stress tensor.

Under the Boussinesq hypothesis, the deviatoric part of the Reynolds stress tensor is expressed in terms of the velocity tensor gradient $S_{ij}=\frac{\partial u_i}{\partial x_j}$ as

\begin{equation}
{R}_{i j}-\frac{1}{3} {R}_{k k} \delta_{i j}=-2 \nu_{t} S_{i j}
\end{equation}

where $\nu_{t}$ is the kinematic eddy viscosity, which requires employing some RANS turbulence model, and $\delta_{i j}$ is the Kronecker delta.

As stated earlier, the turbulence closure selected for the present applications is the Spalart-Allmaras (SA) turbulence model. This model was proposed by \cite{spalart1992one} and solves a single transport equation for a {\it modified} kinematic turbulent eddy viscosity $\tilde{\nu}$. This parameter is identical to $\nu_t$ except in the near-wall region affected by viscosity.

\begin{equation}
\nu_{t}=\tilde{\nu}\left(\frac{\chi^{3}}{\chi^{3}+C_{v 1}^{3}}\right)
\end{equation}

The transport equation for $\tilde{\nu}$ is
\begin{equation}
\begin{split}
\frac{\partial \tilde{\nu}}{\partial t}+
u_{j}\frac{\partial\tilde{\nu}}{\partial x_{j}} =
\frac{1}{\sigma}\left[\frac{\partial}{\partial x_{j}}\left((\nu+ \tilde{\nu}) \frac{\partial \tilde{\nu}}{\partial x_{j}}\right)+C_{b2}\frac{\partial \tilde{\nu}}{\partial x_{i}}\frac{\partial \tilde{\nu}}{\partial x_{i}}\right]+
C_{b 1}(1-f_{t2})\tilde{S}\tilde{\nu}-\\
\left[C_{w1}f_{w}-\frac{C_{b1}}{\kappa^2}f_{t2}\right]\left(\frac{\tilde{\nu}}{d}\right)^2
\end{split}
\end{equation}

where the various parameters are defined as
\begin{equation}
\begin{array}{l}
f_{\omega} = g\left[\frac{1+C_{\omega 3}^{6}}{g^{6}+C_{\omega 3}^{6}}\right]^{1 / 6},\quad g = r + C_{\omega 2}(r^6-r), \quad r = \min\left[\frac{\tilde{\nu}}{\tilde{S}\kappa^2d^2},10\right] \\
\tilde{S} = \Omega + \frac{\tilde{\nu}}{\kappa^2d^2}f_{v2}, \quad f_{t2}=C_{t3}\exp(-C_{t4}\chi ^2), \quad f_{v2} = 1-\frac{\chi}{1+\chi f_{v1}} \\
f_{v1} = \frac{\chi^3}{\chi^3+C_{v1}^3}, \quad \chi = \frac{\tilde{\nu}}{\nu}
\end{array}
\end{equation}

The parameter $d$ is the distance from any given point to the nearest wall, and $\Omega$ is the magnitude of the vorticity. The different constants needed by the model take the following default values

\begin{equation}
\begin{array}{l}
C_{b 1}=0.1355,\quad C_{b 2}=0.622,\quad \sigma=\frac{2}{3},\quad C_{\omega 1}=\frac{C_{b 1}}{\kappa^{2}}+\frac{\left(1+C_{b 2}\right)}{\sigma} \\
C_{\omega 2}=0.3,\quad C_{\omega 3}=2.0,\quad {C_{v1}}=7.1, \quad \kappa = 0.4187,\quad C_{t3}=1.2 \\
C_{t4}=0.5
\end{array}
\end{equation}

\section{Domain, boundary conditions and mesh} \label{sec:numericalModelling}

\begin{figure}[h!]
  \centering
  \hspace{-1.4cm}
  \subfloat[]{\label{fig:domain_all3}
    \includegraphics[width=0.45\textwidth]{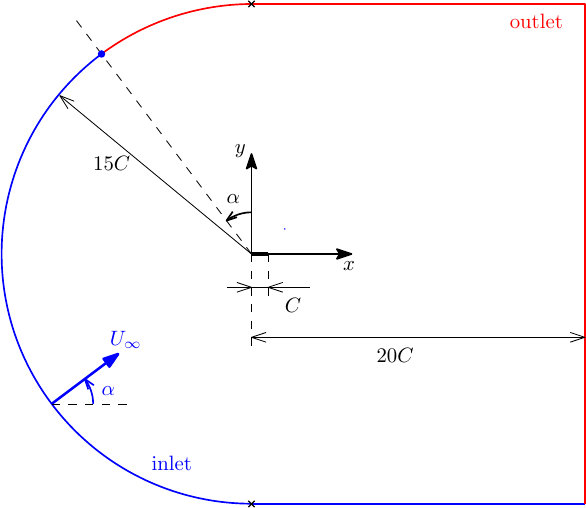}}
  \subfloat[]{\label{fig:domain_jet3}
    \includegraphics[width=0.45\textwidth, trim= 0 0 0 3cm, clip]{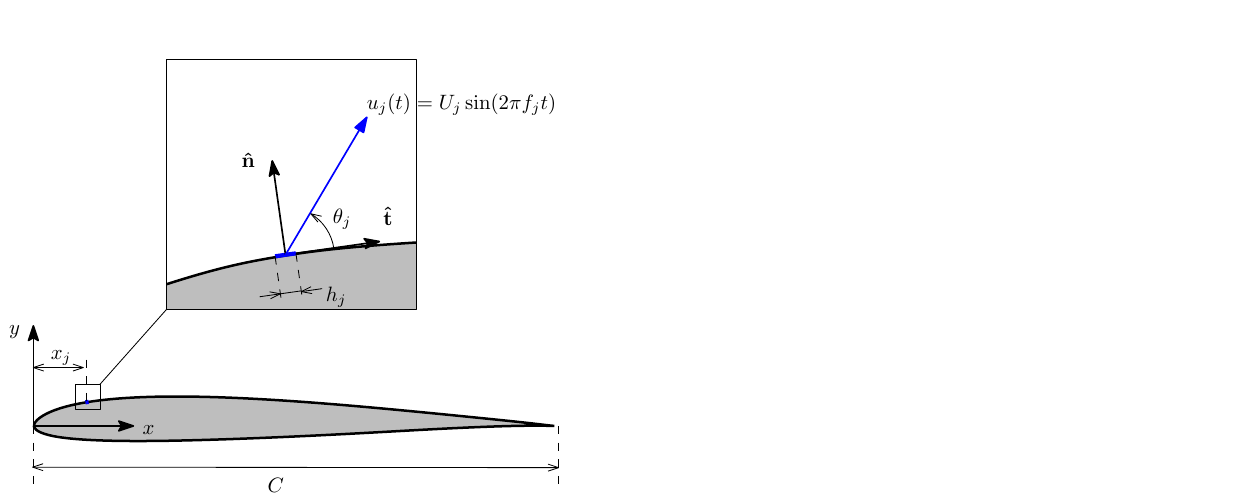}}	
  \caption{(a) Computational domain. (b) Synthetic jet main view.}\label{fig:domain3}
\end{figure}

Fig.~\ref{fig:domain_all3} shows the computational domain around a SD7003 airfoil with a chord length of $C=1$. The leading edge is located at the origin of the coordinate system and the horizontal distance between this point and the inlet to the computational domain (blue curve) is set to $15C$. The outlet of the computational domain (red line) is located $19C$ downstream from the trailing edge of the airfoil. The airfoil is drawn horizontally independently of the simulated angle of attack, which will instead be prescribed by tilting the free-stream velocity. For the cases where the SJA is implemented, the upper surface is split in three at the jet location, with the middle section representing the groove through which the fluid is injected/sucked. Fig.~\ref{fig:domain_jet3} depicts the airfoil with a generic synthetic jet implementation on the upper surface nearby the leading edge. The geometrical design parameters of the synthetic jet (jet angle $\theta_j$, position $x_j$ and width $h_j$), are shown in the zoomed view.

The boundary conditions (BC) used for the velocity and pressure are as follows. At the computational domain inlet, a constant velocity profile is imposed; its components are set to $(u, v) \equiv (U_{\infty}\cos\alpha, U_{\infty}\sin\alpha)$ where $\alpha$ is the angle of attack. Zero-gradient Neumann BC are prescribed for the inlet pressure. At the outlet, Dirichlet BC for pressure and zero-gradient Neumann for velocity are enforced. On the airfoil surface (wall), no-slip condition for velocity and Neumann for pressure is applied.
For the active flow control cases, the jet BC is specified by a time-varying uniform velocity and a zero-gradient Neumann BC for pressure. Specifically, the instantaneous jet velocity is given by $u_j = U_j \sin(2 \pi f_j t)$, where $U_j$ is the maximum jet velocity and $f_j$ is the frequency of the jet oscillation. A top-hat velocity profile is chosen for the SJA, as already employed in \cite{Tousi2021,traficante2016flow,zhang2019comparison}. 

The SA model solves a single turbulence-related unknown, the modified eddy viscosity $\tilde{\nu}$, for which boundary conditions must be defined at all boundaries. \citet{tousi2022large} observed that actuated cases at $\alpha = 14^{\circ}$ were heavily influenced by the free-stream turbulence levels prescribed at domain inlet. In order to investigate this relationship, we initially decided to conduct a parametric study by exploring a wide range of $\tilde{\nu}/\nu\ \in [10^{-20},10^{2}]$ at the inlet while employing Neumann boundary conditions at the jet and outlet boundaries. The BC imposed at the wall was homogeneous Dirichlet $\tilde{\nu}= 0$, and, therefore, no wall function was used.

The aim of the present paper is to further assess the impact of inlet boundary values of $\tilde{\nu}$ at both the jet and domain inlet boundaries. To simplify the discussion, we have labeled these combinations as DN, DD, NN and ND, where the first letter refers to the domain inlet and the second to the jet inlet, 
and the letter itself denotes the type of boundary condition, namely Dirichlet (D) or Neumann (N). It is worth noting that the DN configuration was already analyzed by \citet{tousi2022large} and those results are used for comparison here. 

\begin{figure}[H]
  \centering
    \includegraphics[width=1\textwidth]{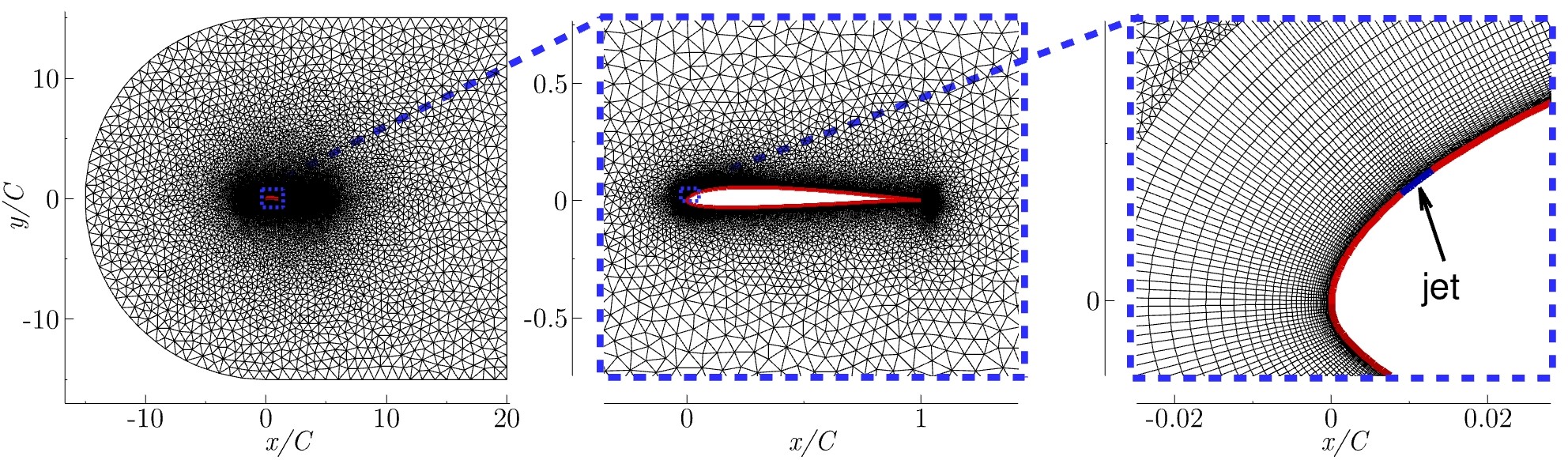}\label{}
  \caption{Mesh overview and zoomed views around the airfoil and leading edge}\label{fig:mesh3}
\end{figure}

The mesh convergence analysis was performed in \citet{Tousi2021}, concluding that a hybrid grid consisting of 45466 cells and $y^{+}$=0.3 was amply sufficient to obtain mesh-independent results. General and detailed views of the mesh are shown in Fig.~\ref{fig:mesh3}.
The near-wall region ($0.04C$-thick) has been meshed with structured quadrilateral elements aligned with a hypothetically attached boundary layer for enhanced accuracy. Further out, the mesh transitions to unstructured triangular cells that grow progressively away from the airfoil surfaces for computational efficiency. A particularly dense structured mesh has been deployed in the neighbourhood of the synthetic jet. The full sensitivity study has been performed for the same set of optimal SJA parameters computed with LES by \cite{tousi2022large} in an unstructured mesh of 28.7 million cells. The mesh was designed to produce aerodynamic coefficients consistent with those reported in the literature and was fine enough to accurately resolve over 95\% of the turbulent kinetic energy, leaving less than 5\% to be modeled by the sub-grid scale model. Here we use these results as a benchmark for our sensitivity analysis.

\textcolor{black}{\citet{Tousi2021} showed that a time step $\Delta t=2\times10^{-5}$ is small enough to resolve the dynamics of the pulsating jet with sufficient accuracy. In order to obtain statistically-converged solutions, the code was evolved for 20 advective time units past all transients and averaging performed over the following 75 time units to include a sufficient number of vortex-shedding cycles. Each simulation, performed with the OpenFOAM CFD package, required about 130h in an Intel\textsuperscript{\textregistered} Core\texttrademark\ i7-7700HQ CPU @ 2.80GHz $\times$ 8 processor server.}              

\section{Results}\label{sec:results3}

The optimal set of SJA parameter values that maximised aerodynamic efficiency of the SD7003 airfoil at $Re = 60,000$ and $\alpha=14^\circ$ was found by \citet{Tousi2021} to be $(f_j,C_\mu,\theta_j,x_j,h_j)=(2.6 C/U_\infty, 0.0055, 18^\circ, 0.0089C, 0.005C)$ using RANS-SA. Even though a different optimum was reported there for maximum lift coefficient, \citet{tousi2022large} exposed via well-resolved LES that the maximum efficiency case outperformed the alleged maximum lift coefficient case in terms of both performance parameters. Despite the fair prediction of aerodynamic performances enhancement, the mismatch between RANS-SA and LES optima unveiled the shortcomings of the former in capturing the flow physics of boundary layer reattachment. In particular, the RANS-SA outcome of successfully actuated cases showed high sensitivity to the boundary conditions for $\tilde{\nu}$ at inflow boundaries, particularly in terms of flow topology.

In the present study, we focus on the maximum efficiency case and explore all four possible combinations (DD, DN, ND and NN) of $\tilde{\nu}/\nu$ boundary condition types (Dirichlet -D- or Neumann -N) at the computational domain inlet (first letter of the pair) and AFC jet inflow (second letter), as well as a wide range of boundary values. Table~\ref{table1} summarises all cases run.

\begin{table}
\centering
\begin{tabular}{c|c|c|c|}
    \multicolumn{2}{c}{} & \multicolumn{2}{c}{Jet inflow}\\
    \hhline{~~--}
    \multicolumn{2}{c|}{} & Dirichlet & Neumann \\
    \hhline{~---}
    \multirow{6}{*}{\rotatebox[origin=c]{90}{Domain inlet}} & \multirow{3}{*}{Dirichlet} & case DD & case DN \\
    & & $(\nu/\tilde{\nu})_i=10^{-20}$ & $(\nu/\tilde{\nu})_i\in[10^{-20},10^{2}]$ \\
    & & $(\nu/\tilde{\nu})_j\in[10^{-20},10^{2}]$ & $\partial_n(\nu/\tilde{\nu})_j=0$\\
    \hhline{~---}
    & \multirow{3}{*}{Neumann} & case ND & case NN \\
    & & $\partial_n(\nu/\tilde{\nu})_i=0$ & $\partial_n(\nu/\tilde{\nu})_i=0$ \\
    & & $(\nu/\tilde{\nu})_j\in[10^{-20},10^{2}]$ & $\partial_n(\nu/\tilde{\nu})_j=0$ \\
    \hhline{~---}
\end{tabular}
\caption{Summary of cases. Rows and columns correspond to $\nu/\tilde{\nu}$ boundary condition types for domain inlet (subindex $i$) and jet inflow ($j$), respectively. Dirichlet and Neumann types are denoted by D and N, and the values used are specified in the corresponding cell. $\partial_n$ is the derivative normal to the corresponding boundary.}
\label{table1}
\end{table}

All other boundary conditions have been kept the same, along with the set of SJA parameters that maximise aerodynamic efficiency.
For the ND case, Neumann boundary conditions for $\nu/\tilde{\nu}$ at inflows have been taken homogeneous (zero gradient) all along, while Dirichlet boundary conditions have been prescribed within the range $\tilde{\nu}/\nu \in [10^{-20}, 10^{2}]$. For the DD case, however, $\tilde{\nu}/\nu$ has been fixed to $10^{-20}$ at domain inlet and only the jet inflow value has been varied in the aforementioned range.

\begin{figure}
  \centering
  \subfloat[]{\label{fig:nut_variation_A3}
    \includegraphics[width=0.5\textwidth, trim=0cm 0 0cm 0 ,clip]{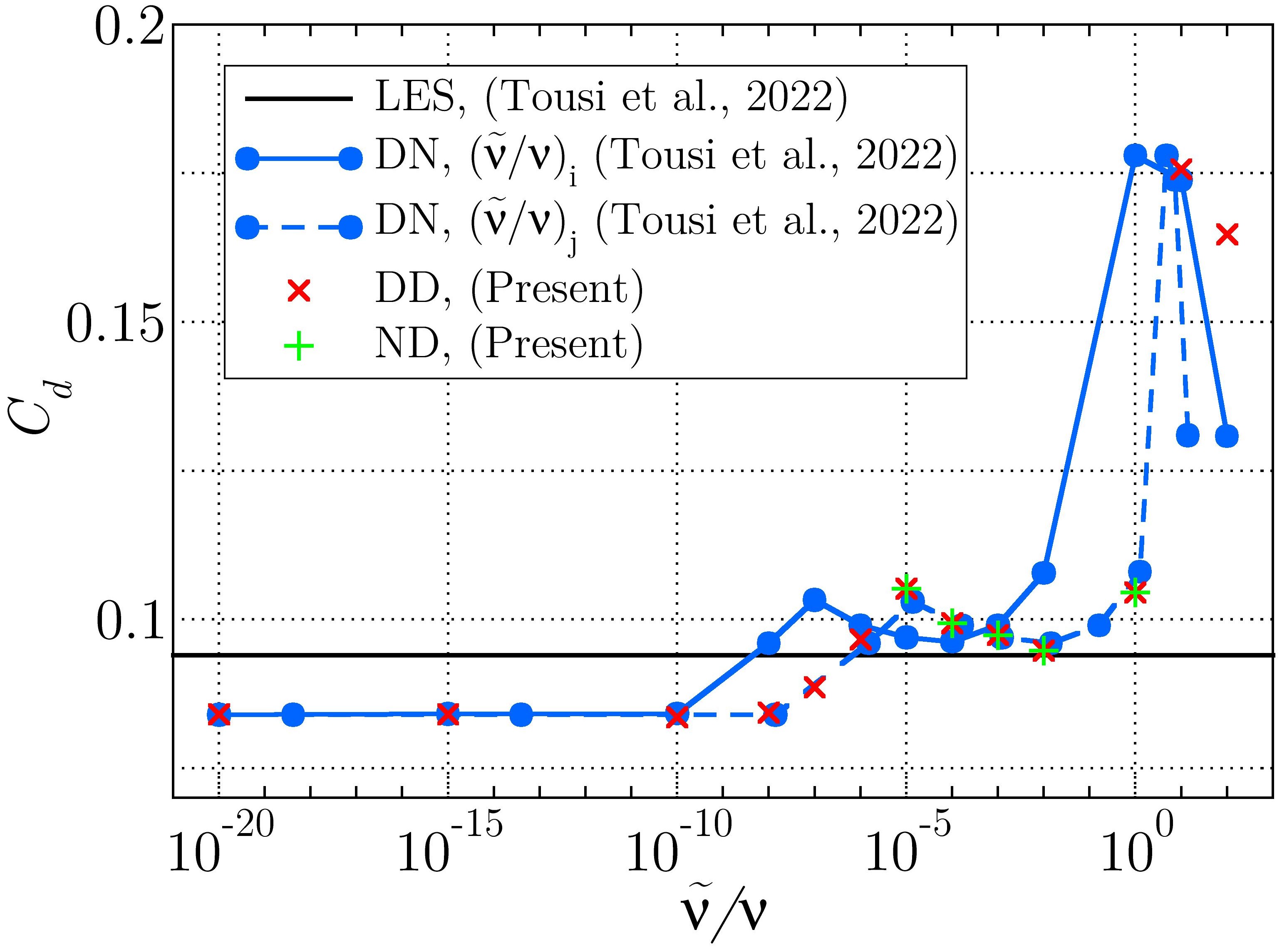}}
  \subfloat[]{\label{fig:nut_variation_B3}
    \includegraphics[width=0.5\textwidth, trim=0cm 0 0 0 ,clip]{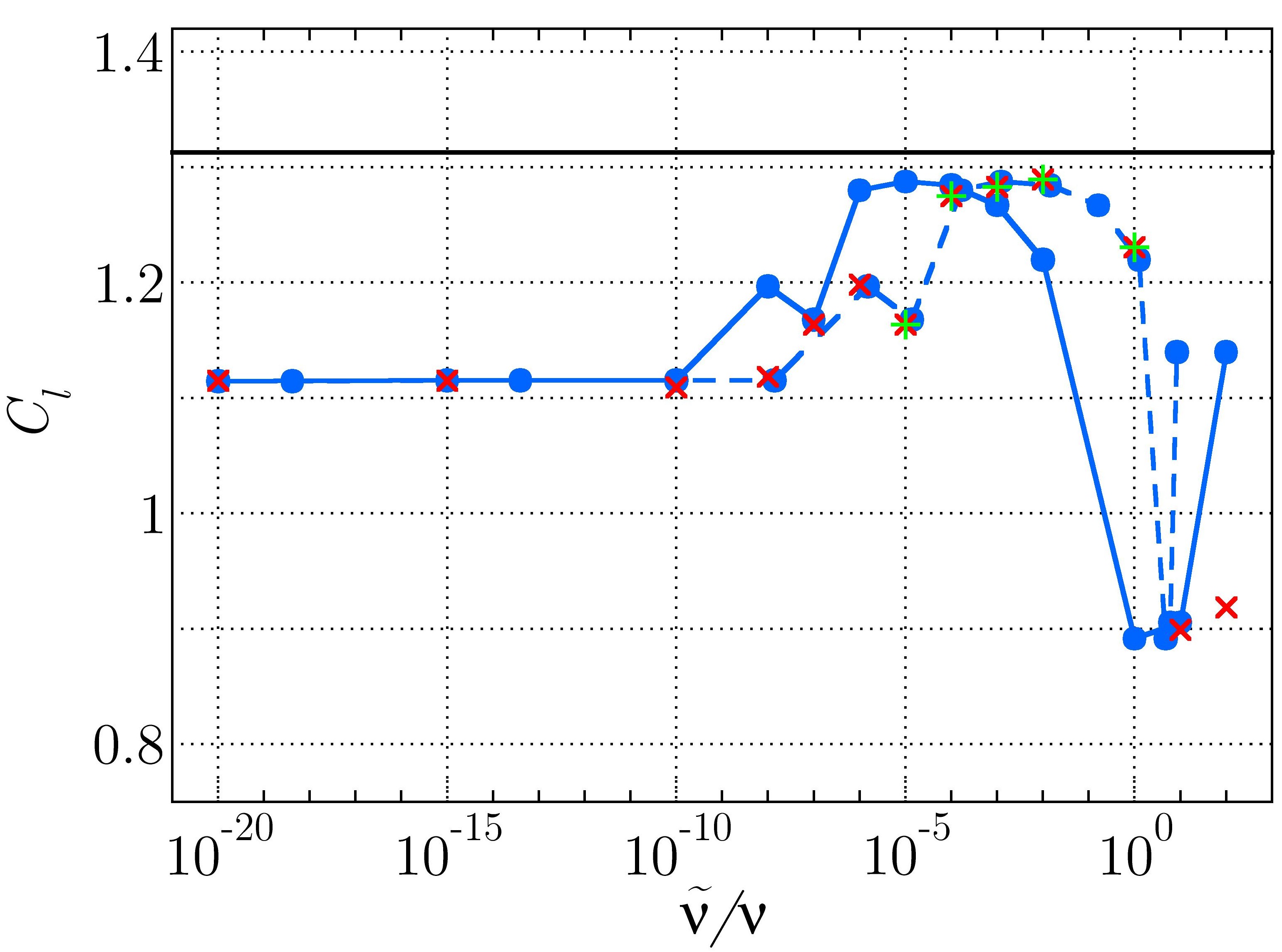}}
  \caption{Sensitivity of time-averaged (a) lift $C_l$ and (b) drag $C_d$ coefficients to inflow preturbulence levels. Shown are the results for case DN taken from \citet{tousi2022large}, first in terms of prescribed $(\tilde{\nu}/\nu)_i$ at domain inlet (solid blue) and then of resulting $(\tilde{\nu}/\nu)_j$ at jet inflow, case DD (red) and case ND (green). The horizontal black line indicates LES results.}
  \label{fig:nut_variation3}
\end{figure}

Fig.~\ref{fig:nut_variation3} shows time-averages of $C_d$ and $C_l$ as a function of $\tilde{\nu_i}/\nu$ at the Dirichlet inflow. The DN cases of \citet{tousi2022large} is shown in two different ways. First, as a function of the value $(\tilde{\nu}/\nu)_i$ prescribed at the inlet of the domain (solid blue line). Then, as a function of the value $(\tilde{\nu}/\nu)_j$ resulting at jet inflow (blue dashed). For low values of $\tilde{\nu}/\nu$, both $C_d$ and $C_l$ are underestimated with respect to LES (black horizontal line, taken from \citet{tousi2022large}). For high values, $C_l$ is again under-predicted while $C_d$ is notably overestimated. In between, there is a range of $(\tilde{\nu}/\nu)_j\in [10^{-4},10^{-1}]$ for which $C_l$ and $C_d$ estimates seem quite accurate. The second curve is shifted as the values of the eddy viscosity are higher at the jet than they were at domain inlet. This results from the boundary layer acting as an amplifier of turbulence due to a high receptivity.
 The first thing that draws attention is the overlapping of the DN (reinterpreted in terms of jet inflow turbulence, blue dashed line) and DD (red crosses) cases, which indicates that the jet turbulence level is key in reproducing the right dynamics. It is also noteworthy that the effect is the same irrespective of whether the turbulent excitation of the jet is directly prescribed or arises from the amplification of free-stream turbulence as a result of boundary layer receptivity. 
For $(\tilde{\nu}/\nu)_j > 10^{-2}$) the airfoil behaves as if no actuation were applied (baseline case). Excessive values of $\tilde{\nu}$ at the jet inflow overrules the eddies originating from the actuation frequency, which consequently leads to a reduction in the momentum exchange efficiency between the jet and the separated boundary layer.

\textcolor{black}{As expected, applying Neumann boundary conditions at both the inlet and jet boundaries generates an indeterminacy that results in unrealistic flow fields. This is the reason why the (NN) case is not analysed.} Failure to prescribe an input value at some of the inflow boundaries leaves the outcome dependent on the initial value of $\tilde{\nu}/\nu$ across the domain at the start of the simulation.

Deactivating the Dirichlet boundary condition at domain inlet (case ND, green plus signs) leaves the results unaltered in comparison with DN and DD. This is further evidence that a realistic jet turbulence input is instrumental in capturing the reattachment dynamics of boundary layers in fluidic AFC applications. While an adequate free-stream pre-turbulence level might be necessary to simulate the unactuated baseline case, its role becomes irrelevant as jet pre-turbulence pervades the boundary layer and dictates its behaviour.

In the remainder of the manuscript, three specific RANS-SA cases have been chosen from the parametric analysis just discussed in order to assess the impact of the inflow turbulence level on aerodynamic performance and flow topology. Two DD cases with $(\tilde{\nu}/\nu)_j = 10^{-3}$ and $10^{-5}$ at the jet inflow have been chosen and named D$_{20}$D$_{3}$ and D$_{20}$D$_{5}$, respectively. Besides LES, these have also been compared against a third case D$_{5}$N from \citet{tousi2022large} with $(\tilde{\nu}/\nu)_i = 10^{-5}$ and a fourth ND$_{3}$ with $(\tilde{\nu}/\nu)_j = 10^{-3}$.
Table~\ref{tab:perf} summarises the results.

\begin{table}[h!]
\centering
\begin{tabular}{@{}llccccc@{}}
\toprule
Case & Turb. model & \multicolumn{2}{c}{$\tilde{\nu}/\nu$ or $\partial_n(\tilde{\nu}/\nu)$}  & $C_l$ & $C_d$ & $\eta=\dfrac{C_l}{C_d}$ \\ \cmidrule(lr){3-4}
\multicolumn{2}{c}{} & inlet & jet & \multicolumn{3}{c}{} \\ \midrule
LES \cite{tousi2022large} & LES & - & - & 1.313 & 0.094 & 13.96 \\
D$_{5}$N \cite{tousi2022large}  & \multirow{3}{*}{RANS-SA} & $10^{-5}$ & 0 & 1.288 & 0.097 & 13.27 \\
D$_{20}$D$_3$ & & $10^{-20}$ & $10^{-3}$ & 1.283 & 0.097 & 13.17 \\
D$_{20}$D$_5$ &  & $10^{-20}$ & $10^{-5}$ & 1.164 & 0.105 & 11.06 \\
ND$_3$ & & 0 & $10^{-3}$ & 1.283 & 0.097 & 13.17  \\ \bottomrule
\end{tabular}%
\caption{Aerodynamic coefficients for LES and several RANS-SA cases.}
\label{tab:perf}
\end{table}

As was already clear from Fig.~\ref{fig:nut_variation3}, the cases D$_{5}$N, D$_{20}$D$_3$ and ND$_3$ produce very similar aerodynamic performances, in fair agreement to LES results. They share the same (or similar for the DN case) values of $\tilde{\nu}/\nu$ at the jet inflow boundary. The D$_{20}$D$_5$ case underestimates aerodynamic performances, presumably on account of too low pre-turbulence levels at the jet inflow.

\begin{figure}
  \centering
  \subfloat[]{\label{fig:cp}
    \includegraphics[width=0.45\textwidth, trim=0cm 0 0cm 0 ,clip]{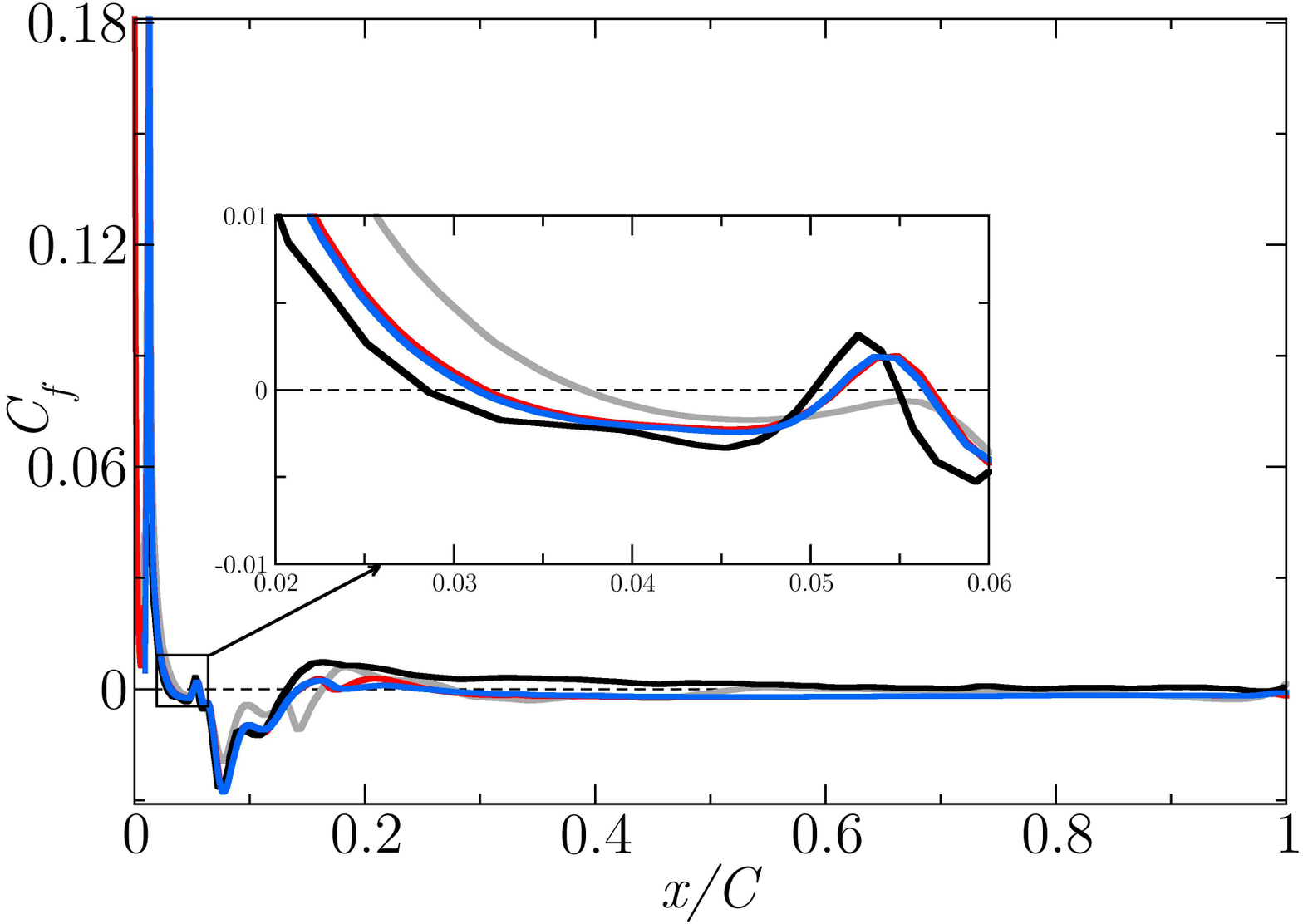}}
  \subfloat[]{\label{fig:cf}
    \includegraphics[width=0.45\textwidth, trim=0cm 0 0 0 ,clip]{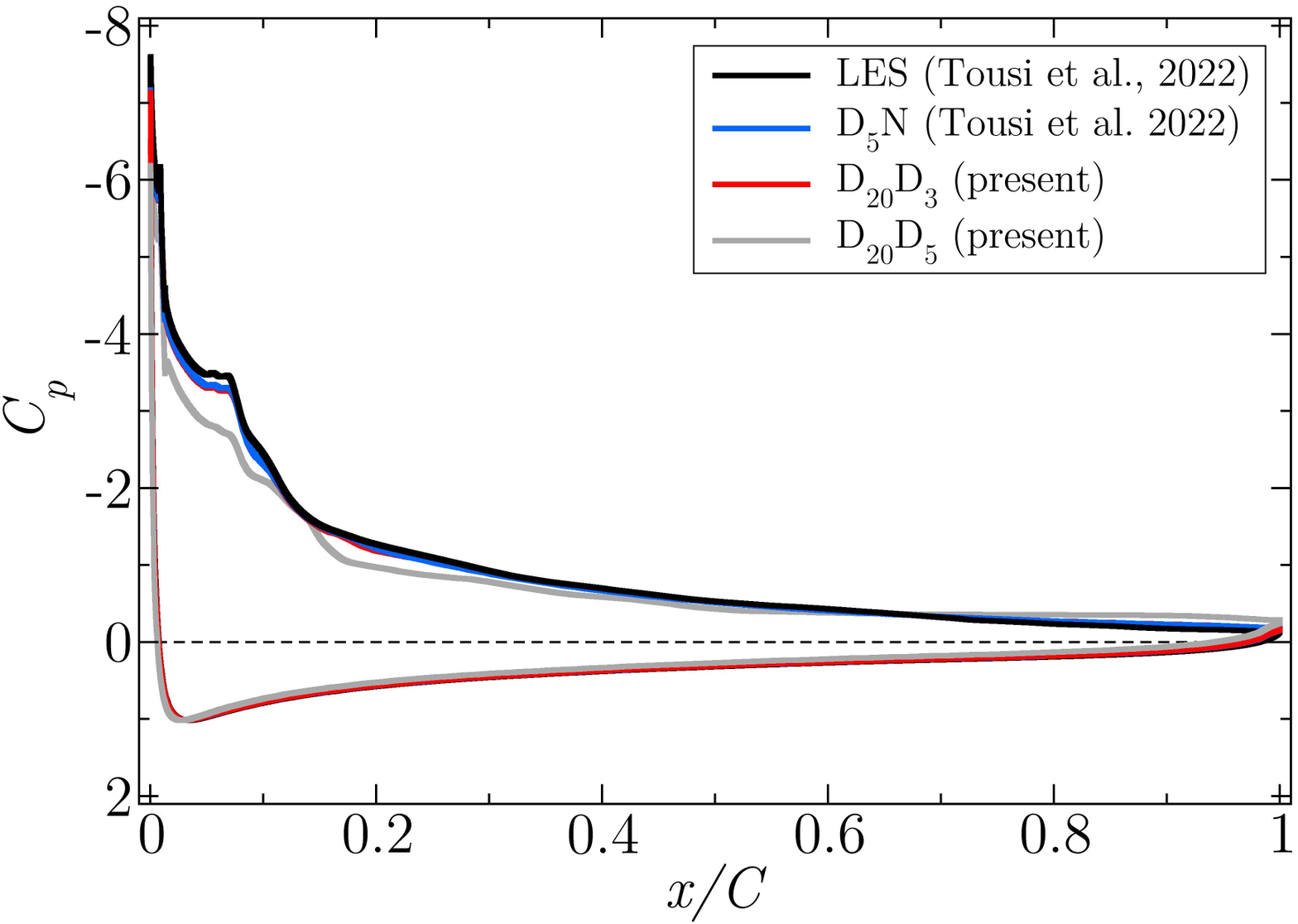}}
  \caption{Chord distribution of the (a) Skin friction $C_f$ and (b) pressure $C_p$ coefficients for selected cases of table~\ref{tab:perf}. The inset helps identify separation for the various cases.
  }
  \label{fig:cpcf}
\end{figure}

Fig.~\ref{fig:cpcf} shows chord distributions of time-averaged pressure $C_p = 2(p-p_{\infty}) / \rho {u^{2}_\infty}$ and skin-friction $C_f = 2 \tau_w / \rho {u^{2}_\infty}$ coefficients for the D$_{20}$D$_3$ (red), D$_{20}$D$_5$ (gray), D$_{5}$N (blue) and LES (black) cases. Only the distributions along the upper surface are shown for $C_f$. The ND$_3$ case is indistinguishable from D$_{20}$D$_3$ and is, accordingly, not shown.  

Cases D$_{20}$D$_3$ and D$_{5}$N aptly reproduce the $C_f$ distribution of LES along the front part of the airfoil. The separation ($x_s\simeq0.03$) and reattachment ($x_r\simeq0.13$) points are in fair agreement (see the inset) and only beyond the latter point, after the recirculation bubble, do the distributions stray yet, as we will see, with little consequence for the pressure distributions. Too low a turbulence level at the jet, as for case D$_{20}$D$_5$, has the initial separation occur slightly later but, at the same time, the recirculation bubble gets elongated and reattachment is postponed in a way that enhances the mismatch with respect to LES. While LES predicts a fully attached boundary layer after the separation bubble, RANS computations have most of the aft-upper-surface boundary layer detached. Nonetheless, the separation is only mild for the D$_{20}$D$_3$ and D$_{5}$N cases, so that average streamlines just outside the detached boundary layer are barely deviated, and the pressure distribution on the suction side of the airfoil compares favourably with that of LES computations. Only around the jet location, where the laminar-turbulent transition occurs, do these RANS-SA cases slightly underestimate the suction predicted by LES. For the D$_{20}$D$_5$ case, the larger separation considerably bubble reduces the suction over the front half of the airfoil. The somewhat stronger suction in the rear third does not compensate and the mean $C_l$ falls short from LES values. The pressure distribution along the lower surface is indistinguishable among the various cases considered and LES.

\begin{figure}
\begin{tabular}{cc}
(a) & (b) \\
\includegraphics[width=0.48\textwidth, trim=0cm 0 0 0 ,clip]{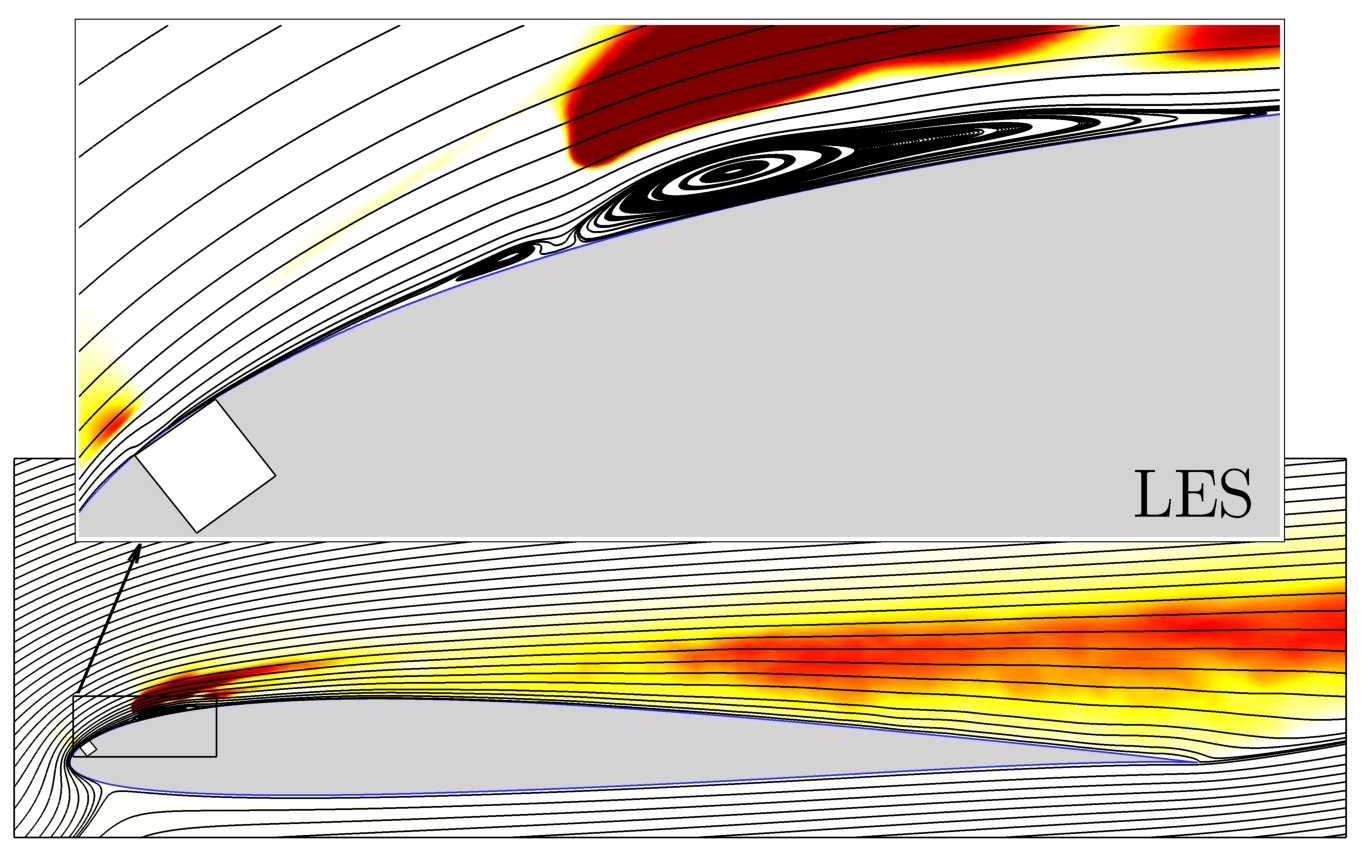} &
\includegraphics[width=0.48\textwidth, trim=0cm 0 0 0 ,clip]{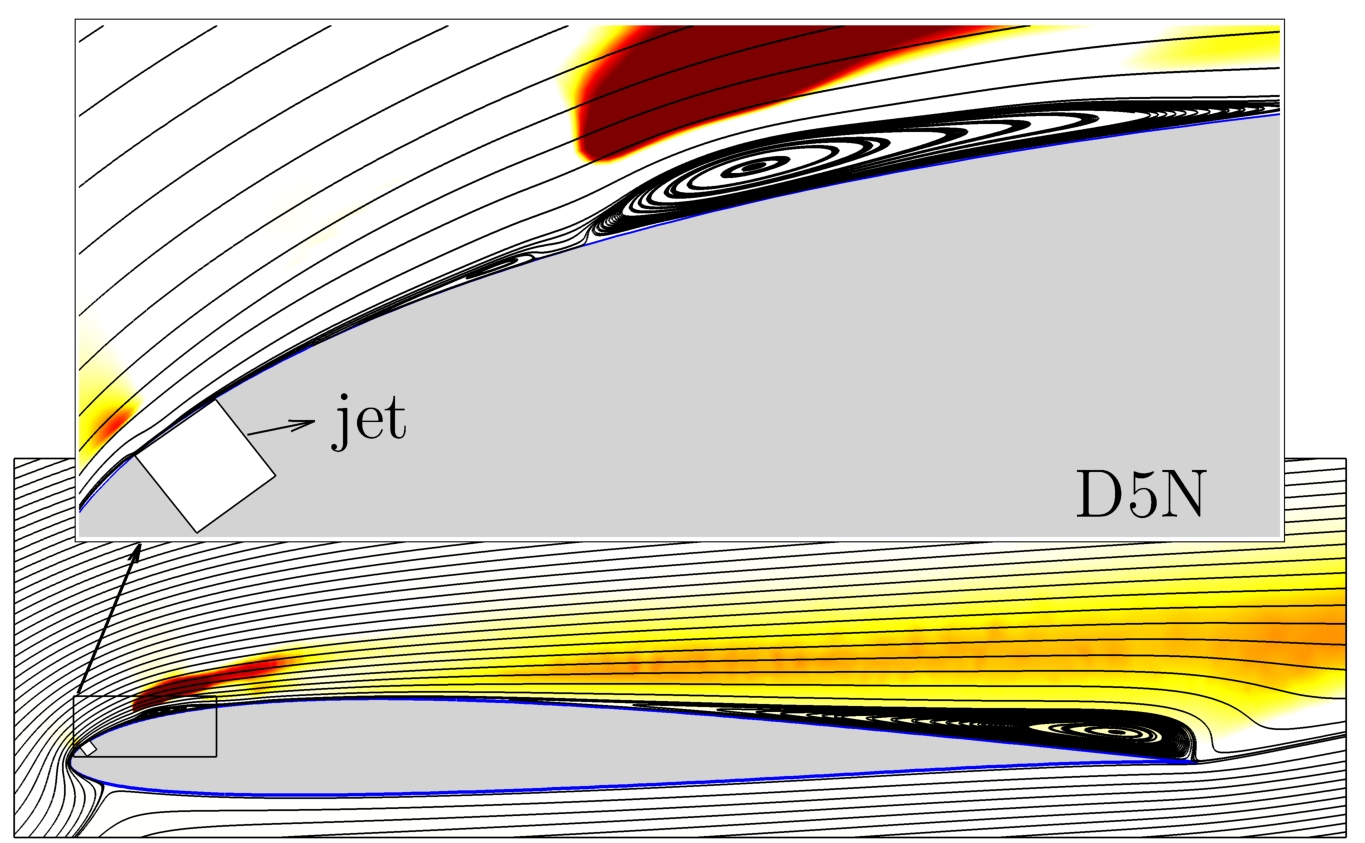} \\
(c) & (d) \\
\includegraphics[width=0.48\textwidth, trim=0cm 0 0 0 ,clip]{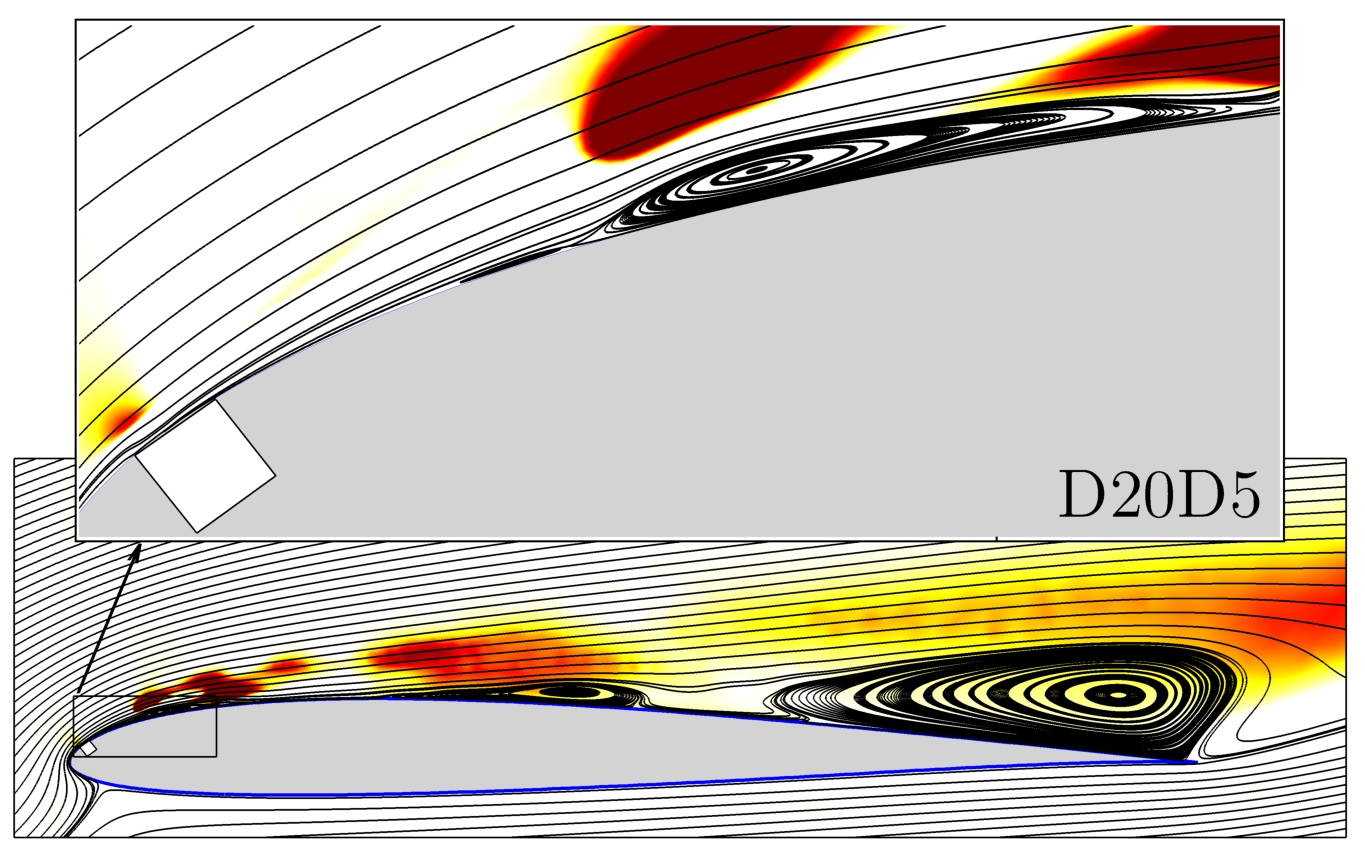} &
\includegraphics[width=0.48\textwidth, trim=0cm 0 0 0 ,clip]{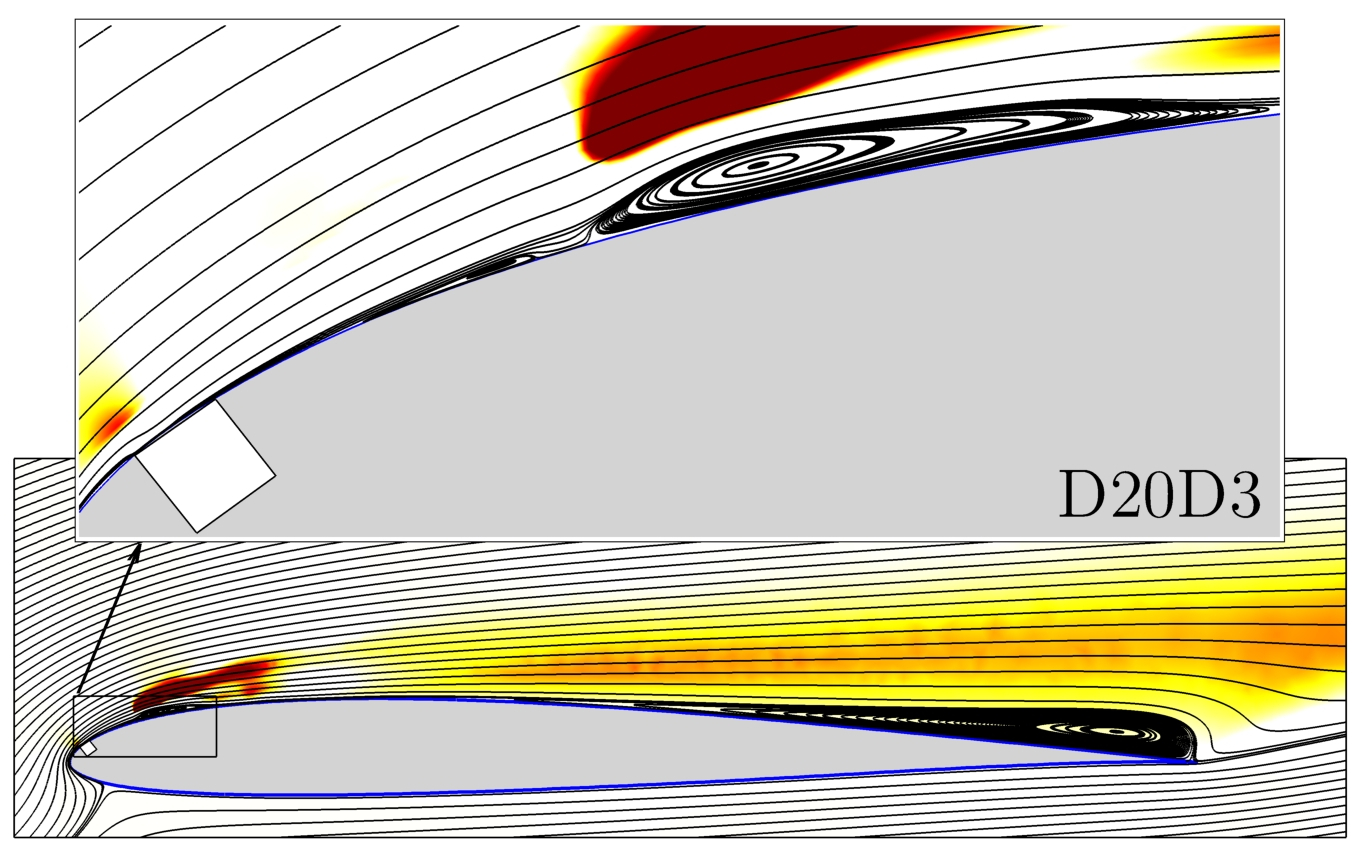} \\
\multicolumn{2}{c}{\includegraphics[width=0.4\textwidth, trim=0cm 0cm 0cm 0 ,clip]{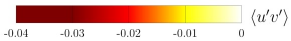}}
\end{tabular}
  \caption{Mean flow topology of cases (a) LES, (b) D$_5$N, (c) D$_{20}$D$_5$ and (d) D$_{20}$D$_3$. Shown are contours of mean streamlines and colourmap of the $\langle u'v'\rangle$ component of the Reynolds stress tensor. Fields have been time averaged and, in the case of LES, also spanwise-averaged.
  }
  \label{fig:Restress}
\end{figure}

Time-averaged streamlines of the various cases listed in table~\ref{tab:perf} are depicted in Fig.~\ref{fig:Restress}, together with colormaps of the $\langle u'v'\rangle$ component of the Reynolds stress tensor. 
For LES, the fields have been also averaged in the spanwise direction. The mean streamlines of the LES computation (Fig.~\ref{fig:Restress}a) that are nearest to the surfaces closely follow the shape of the airfoil. Only very close to the upper surface do the adjacent streamlines detach for a while on the front part and perhaps also at the rear third, but the separation is very mild. Case D$_{20}$D$_5$ (Fig.~\ref{fig:Restress}c) shows instead a long separation bubble at mid chord and a massive recirculated region at the aft third of the chord. Close inspection also reveals that the front recirculation bubble trailing from the jet location is slightly longer than for LES. Most of these unrealistic effects are corrected with the higher turbulence levels at the front portion of the boundary layers of cases D$_5$N (Fig.~\ref{fig:Restress}b) and D$_{20}$D$_3$
(Fig.~\ref{fig:Restress}d). The front separation bubble recovers the size predicted by LES and, although a long recirculated region remains on the rear half of the airfoil, its thickness is greatly reduced such that external streamlines are less deviated and pressure resulting distributions better suited to produce lift.

Turbulent levels are evidenced in Fig.~\ref{fig:Restress} through the only statistically relevant shear component of the Reynolds stress tensor $\langle u'v'\rangle$.  There is, in all cases, a strongly turbulent region dowwnstream from the jet location, just slightly milder for D$_{20}$D$_5$. Turbulence is however temporarilly tamed upon boundary layer reattachment and does not show up in the outer flow until later, from mid-chord downstream. This wake turbulence is somewhat underrated by D$_5$N and D$_{20}$D$_3$. D$_{20}$D$_5$ shows a very different pattern, with high turbulence intensity over the mid separation bubble and then strongly boosted anew downstream by the strong recirculation region at the back. It appears that insufficient turbulent levels in the separated boundary layer delay reattachment and induce an early second separation.

\begin{figure}
  \centering
    \includegraphics[width=1\textwidth]{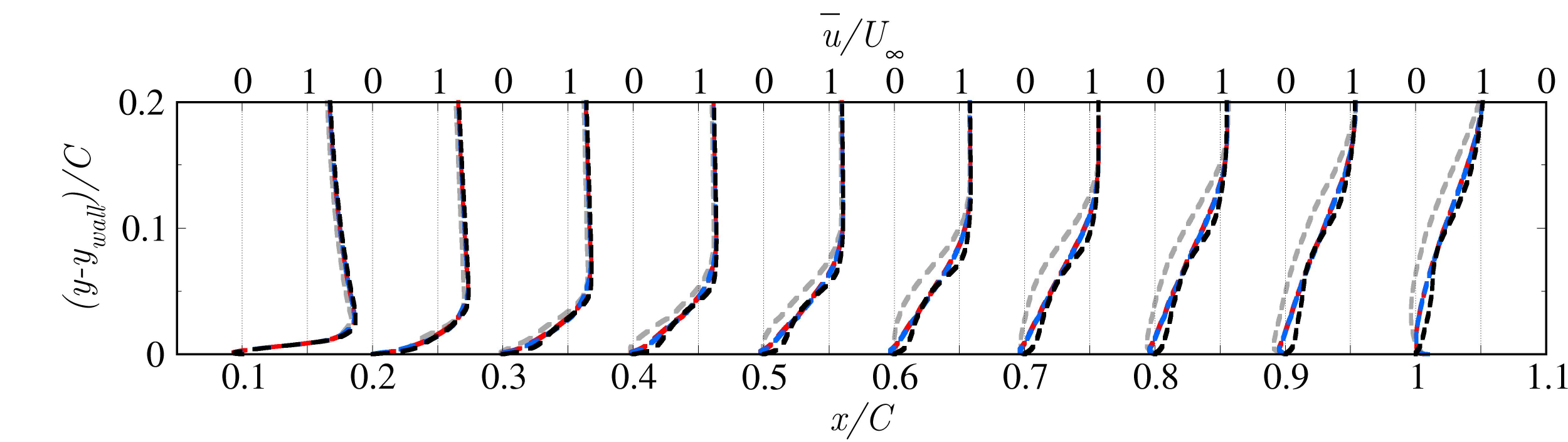}
  \caption{Evolution of the boundary layer developing on the upper surface of the airfoil. Time- and spanwise-averaged (for LES) wall-normal profiles of tangential velocity $\langle u(x,y,z;t)\rangle_{t,z}$ at streamwise coordinates $x/C=0.1$ through $x/C=1$. Shown are the LES (black dashed line), D$_5$N (blue line), D$_{20}$D$_3$ (red line) and D$_{20}$D$_5$ (gray dashed line) cases.
    } \label{fig:BL}
\end{figure}

The evolution of the time-averaged (also spanwise-averaged for LES) wall-normal profiles of wall parallel velocity along the chord is shown in Fig.~\ref{fig:BL} at various streamwise locations on the upper surface. 
The flow profiles obtained from LES (black dashed line), indicate that the boundary layer remains attached over most of the upper surface due to the effect of the synthetic jet actuator. The only location where a very small flow separation is observed, visible as a slight backflow region in close proximity of the wall, occurs at $x=0.1$, where Fig.~\ref{fig:Restress} disclosed the formation of a small separation bubble downstream from the jet location. The velocity profiles for the D$_5$N (blue line) and D$_{20}$D$_3$ (red line) cases are indistinguishable from those for LES over the front half of the upper surface and start deviating, albeit only in very close proximity of the wall, along the rear half. In these region, a slight back flow develops, traceable to the long but thin recirculation bubble detected in Fig.~\ref{fig:Restress}b,d that is not present in the LES calculation. The outer flow is very little disturbed, which explains the fair estimation obtained for the various aerodynamic performance parameters. meanwhile, the velocity profiles for the D$_{20}$D$_5$ case (gray dashed line) start diverging early on in the development of the flow and showcase a much thicker boundary layer than predicted by LES and RANS-SA cases with the right amount of pre-turbulence levels.
\begin{figure}
\begin{tabular}{ccc}

\hspace{-0.5cm}
\includegraphics[width=0.357\textwidth, trim=0cm 0cm 5cm 0, clip]{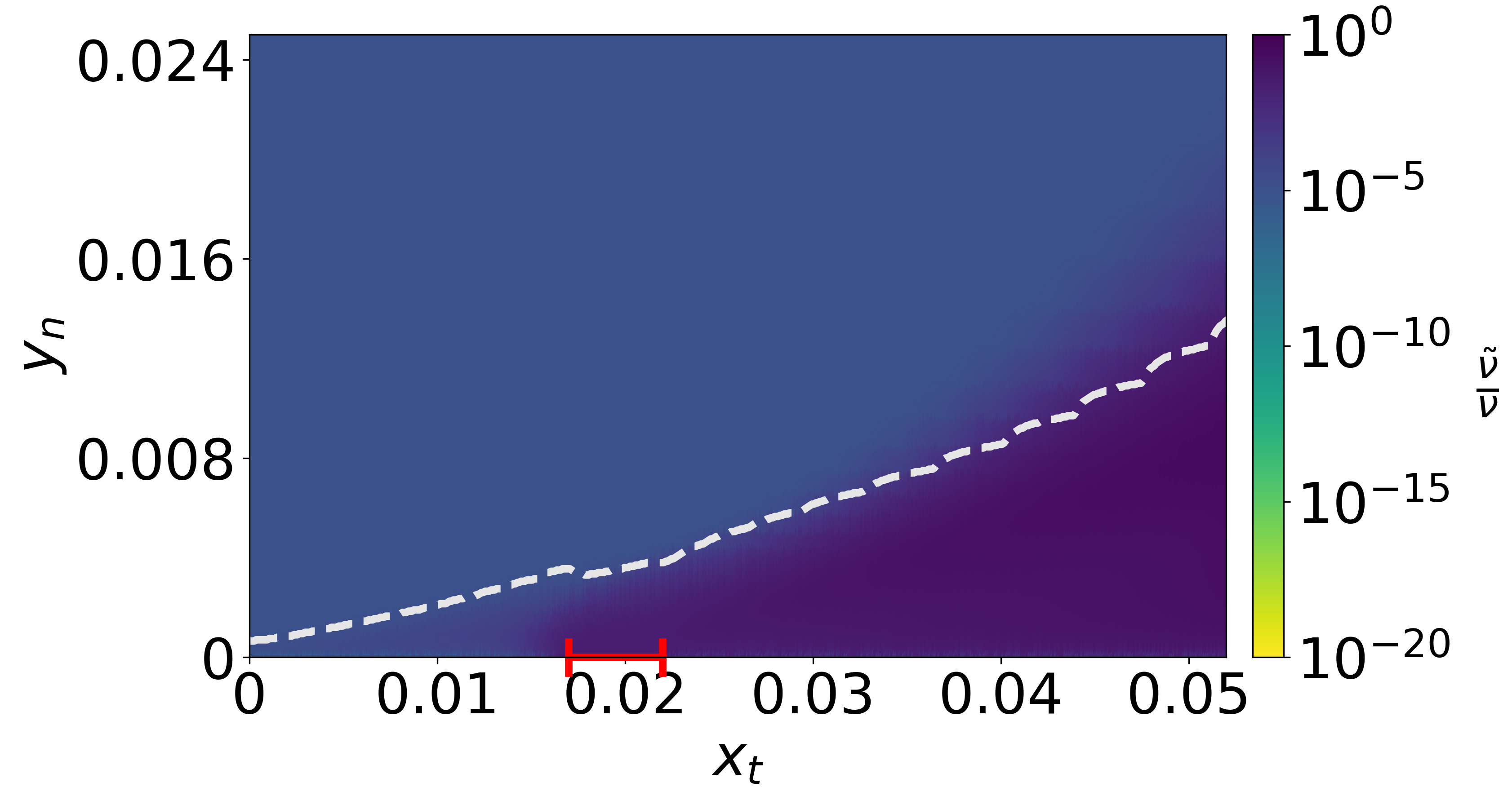} &
\hspace{-0.5cm} %
\includegraphics[width=0.289\textwidth, trim=4.5cm 0 5cm 0, clip]{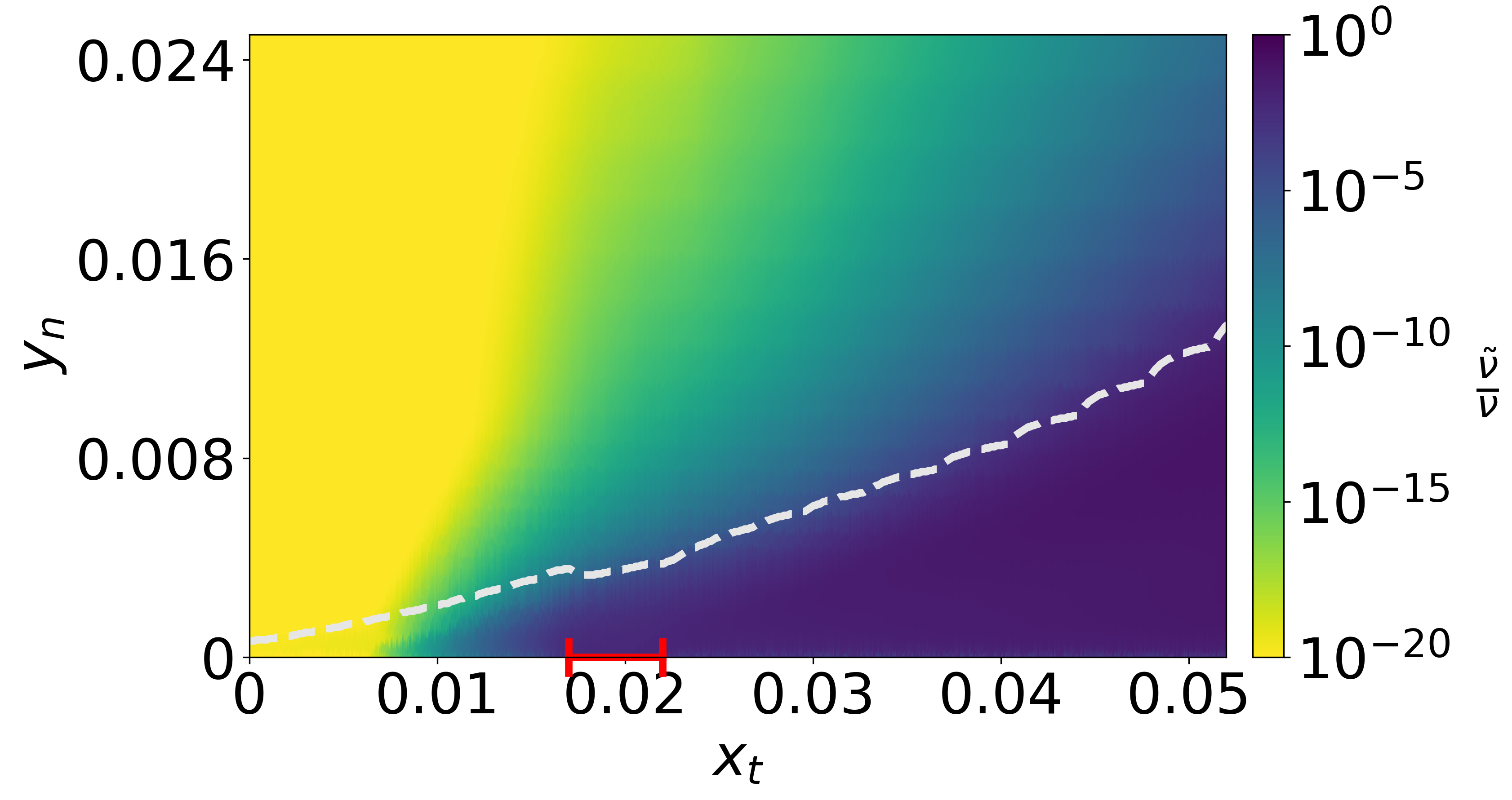} &
\hspace{-0.5cm} %
\includegraphics[width=0.365\textwidth, trim=4.5cm 0 0cm 0, clip]{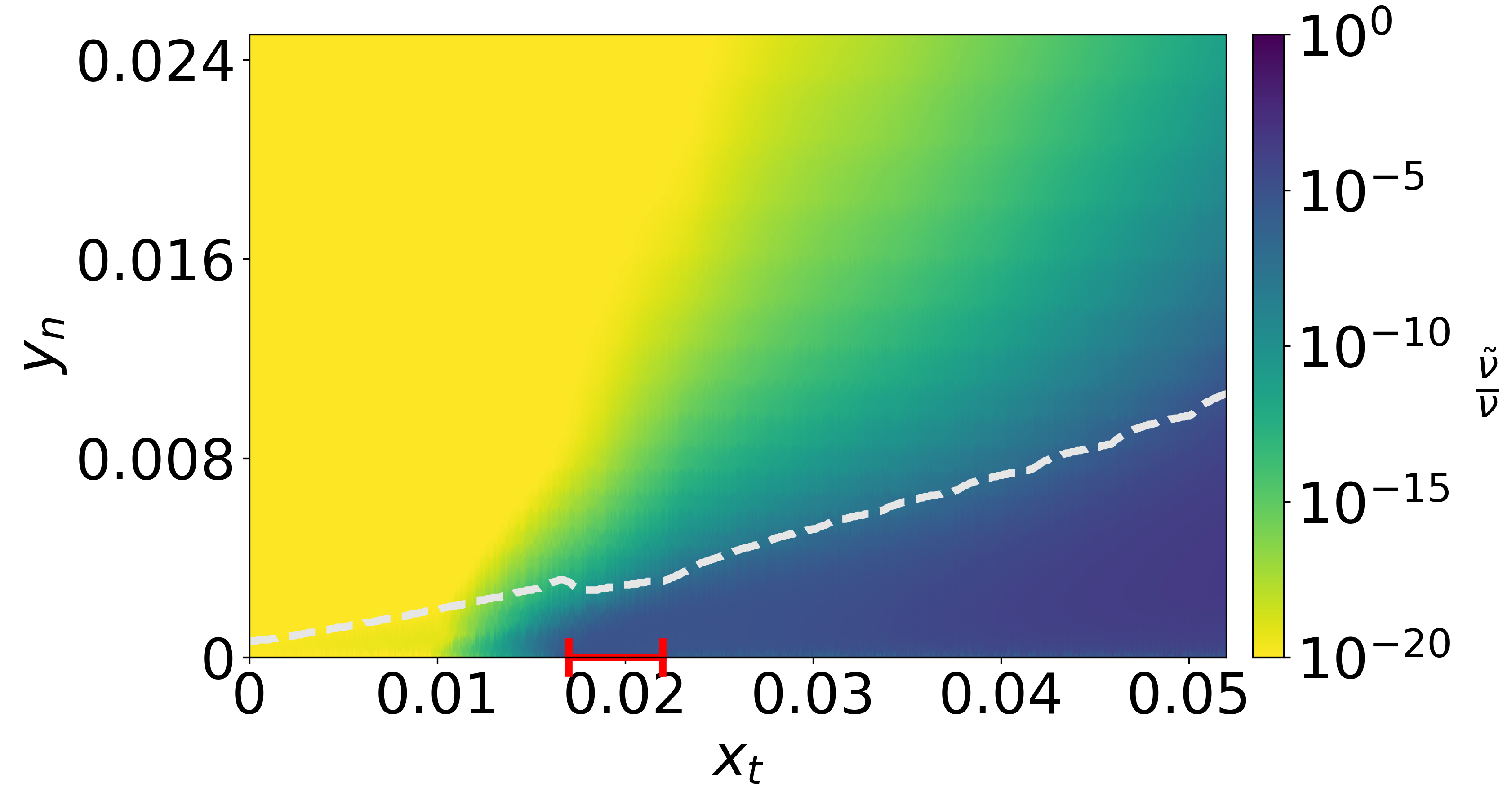} \\
(a) & (b) & (c) \\
\end{tabular}

\caption{Colormaps of $\tilde{\nu}/\nu$ in the foremost part of the boundary layer on the upper surface of airfoil. (a) D$_5$N, (b) D$_{20}$D$_3$, and (c) D$_{20}$D$_5$. The coordinates $x_t$ and $y_n$ measure the arc-length along the upper surface and the wall-normal distance, respectively. The red ticks indicate the position of jet. The white dashed lines delineate the boundary layer threshold, based on 95\% decay of wall shear stress with respect to its value at the wall.}
  \label{fig:nutilda_bl}
\end{figure}

To better visualize and comprehend the mechanism by which $\tilde{\nu}/\nu$ increases from the freestream to the boundary layer, colormaps are shown in Fig.~\ref{fig:nutilda_bl} spanning from the leading edge to past the jet location for the D$_5$N, D$_{20}$D$_{3}$, and D$_{20}$D$_{5}$ cases. The $x_t$ axis measures the local tangential arc-length distance along the upper surface starting at the leading edge while the $y_t$ coordinate is taken locally normal to the surface. The jet position, $(x_t, y_n) = (0.017, 0.022)$ or $(x/C,y/C)=(0.0089,0.0139)$, is indicated by red lines. The boundary layer threshold (white dashed line) is defined as the wall-normal location where the shear stress has diminished to below 5\% of its value at the wall.
A common observation across all three cases is that $\tilde{\nu}/\nu$ free stream levels prescribed at the domain inlet ($10^{-20}$ in D$_{20}$D$_3$ and D$_{20}$D$_5$, and $10^{-5}$ in D$_5$N) are barely altered while in the inviscid region and are instead boosted within the boundary layer. The increase rate of $\tilde{\nu}/\nu$ beyond the jet location is similar in all cases, but since D$_{20}$D$_5$ starts with a lower value, the absolute level is about two orders of magnitude below. D$_5$N and D$_{20}$D$_3$ present instead very similar values of $\tilde{\nu}/\nu$ in the boundary layer everywhere past the jet, but they achieve this in two different ways. While D$_{20}$D$_3$ directly prescribes at the jet an adequate value to be representative of boundary layer conditions (as compared to LES), D$_5$N relies instead on sufficient free stream turbulence that the boundary layer can boost to realistic values upon reaching the jet location. Incidentally, when turbulence is inoculated into the boundary layer from the outside (case D$_5$N), the boundary layer threshold clearly shows that turbulence is being generated inside. If free stream turbulence is negligible (cases D$_{20}$D$_3$ and D$_{20}$D$_3$), the turbulence generated inside the boundary layer diffuses into the inviscid region and is advected downstream by the outer flow.

In accordance, we can conclude that when the RANS-SA turbulence model is employed in SJ-AFC applications, the best agreement with LES computations is obtained when the turbulence level in the boundary layer immediately past the jet location is realistically set to about $\tilde{\nu}/\nu=10^{-3}$. This can be accomplished by prescribing a sufficiently large turbulence level at the domain inlet, as for case D$_5$N, or imposing the adequate turbulence level at the jet inflow, as for case D$_{20}$D$_{3}$. 

\section{Conclusions}\label{sec:conclusions3}

The results presented in this paper highlight the importance of prescribing carefully-tuned and realistic values for the turbulence fields at the inflow boundary conditions when dealing with AFC applications if accurate estimates of macroscopic performance parameters are to be obtained.

Upon optimising a bunch of AFC-SJA parameters for the SD7003 airfoil at a $Re=60,000$ and $14^{\circ}$ angle of attack (post-stall conditions), \citet{Tousi2021, tousi2022large} realised that RANS-SA computations (Reynolds-Averaged Navier-Stokes equations with the Spalart-Allmaras turbulence model) are very sensitive to prescribed turbulence levels at inflow boundaries, particularly so in actuated cases with the jet on. A preliminary exploration revealed that pre-turbulence free stream levels of $\tilde{\nu}/\nu\in[10^{-6},10^{-3}]$ where required to faithfully reproduce the much more accurate LES results.

Here we have systematised the analysis to several combinations of boundary condition types at the free-stream and jet inflow boundaries as well as to a wide range of $\tilde{\nu}/\nu$ values to elucidate what is required from RANS-SA simulations to provide results as accurate as possible both in terms of flow topology and aerodynamic performance indicators.

A first conclusion, perhaps evident, is that a value for $\tilde{\nu}$ must be prescribed at at least one of the inflow boundaries. Applying Neumann boundary conditions everywhere cannot but fail because results are then left to the whim of flow initialistion and numerical method implementation. Our findings show that a value must be set in an inflow Dirichlet boundary such that the boundary layer is sufficiently excited to undergo transition in a similar way as occurs in actual experimental conditions. This can be done at the inlet of the domain, so that the free stream carries the turbulence to the boundary layer threshold, but then the adequate value can depend on domain size.
The jet inflow offers the opportunity to excite the boundary layer with the right amount of turbulent intensity directly. RANS-SA simulations that inject a $\tilde{\nu}/\nu\in[10^{-4},10^{-1}]$ into the incipient boundary layer produce results that align well with LES prediction. In our domain, this can be accomplished with $\tilde{\nu}/\nu\sim10^{-5}$ at the free-stream inlet or $10^{-3}$ directly at the jet inflow. By doing this, not only the lift and drag coefficients are estimated within 3\% of LES prediction, but also the flow fields at large are fairly reproduced, particularly on the front half of the airfoil, where the overall aerodynamic performances are mostly determined.

\backmatter

\section{Funding}

This work was supported by the Spanish and Catalan Governments under grants PID2020-114043GB-I00 and 2021-SGR-00586, respectively. Part of the computations was done in the Red Espa\~{n}ola de Supercomputaci\'{o}n (RES), Spanish supercomputer network, under the grants IM-2019-3-0002 and IM-2020-1-0001.

\bibliography{sn-bibliography02}

\end{document}